\documentclass[reprint]{revtex4-1}
\usepackage{amssymb}
\usepackage{amsfonts}
\usepackage{amsmath}
\usepackage{color}
\include{latin1,german}
\usepackage{inputenc}
\usepackage{epsfig}
\usepackage{float}

\def \be {\begin{equation}}
\def \ee {\end{equation}}

\begin{document}

\date{\today}

\title{Superconducting $\pi$-ring metamaterials}

\author{Derek Michael Forrester}
\affiliation{Department of Chemical Engineering, Loughborough University, Leicestershire, UK, LE11 3TU }

\author{Karl E. K\"{u}rten}
\affiliation{Faculty of Physics, University of Vienna, 5, Boltzmanngasse, A-1090, Vienna, Austria }

\author{Feodor V. Kusmartsev}
\affiliation{Department of Physics, Loughborough University, Leicestershire, UK, LE11 3TU }

\begin{abstract}

We develop the concept of fractal metamaterials which consist of arrays of nano and micron sized rings containing Josephson junctions which play the role of ``atoms'' in such artificial materials. We show that if some of the junctions have $\pi$-shifts in the Josephson phases that the ``atoms'' become magnetic and their arrays can have tuned positive or negative permeabilty. Each individual ``$\pi$-ring'' - the Josephson ring with one  $\pi$-junction - can be in one of two energetically degenerate magnetic states in which the supercurrent flows in the clockwise or counter-clockwise direction. This results in magnetic moments that point downwards or upwards, respectively.  The value of the total magnetization of such a metamaterial may display fractal features. We describe the magnetic properties of such superconducting metamaterials, including the magnetic field distribution in them (i.e. in the network that is made up of these rings). We also describe the way that the magnetic flux penetrates into the Josephson network and how it is strongly dependent on the geometry of the system.

\end{abstract}

\maketitle

\section{Introduction}

Recently it was shown that artificially structuring superconducting rings on the subwavelength scale can produce nonlinear and switchable metamaterials \cite{savinov,Jung,Jung2,Anlage}. Arrays of rings of Josephson junctions or systems of superconducting islands on normal metal films can be excited by electromagnetic radiation \cite{vourdas1997} and support the propagation of electromagnetic waves \cite{everitt}, thus significantly modifying vacuum permeability. Even the sign of the permeability can be changed in such a system. Each superconducting ring is a ``meta-atom'' or ``molecule'' and the arrays of rings make up the ``metamaterial''.  These rings can have currents induced into them by the application of electromagnetic fields. However, special rings containing two superconducting islands that are separated from one another by a weak link (a Josephson junction), generate their own spontaneous current when the difference in the phases of their quantum mechanical wave functions in the ground state energy is $\pi$. These special junctions are known as  $\pi$-junctions and rings containing odd numbers of them are  $\pi$-rings \cite{kusmartsev1992, Kirtley,Ryazanov,Guichard,blatter,Andreev}. The spontaneous currents that circulate in these rings occur when no external current or magnetic field is applied and they are randomly orientated, i.e. clockwise or counter-clockwise. Metamaterials composed of these rings exploit the energy competition resulting from the spontaneous currents and magnetic flux quantization, thus creating a complicated switching distribution. Superconducting metamaterials are highly desirable due to their low ohmic losses, controllability by electromagnetic fields, sensitivity to temperature, and their potential for use in plasmonics. The harnessing of light by metamaterials is limited by dissipative losses. Superconducting metamaterials offer a solution by being inherently low loss devices with negative effective permeability \cite{Zheludev}. Spontaneous magnetization is a particularly intriguing attribute for a metamaterial as it reduces the need for a lot of extraneous circuitry. It also means that the metamaterial structure is already in a magnetic state and that small localized perturbations can be used at one part of the metamaterial to affect the whole structure, much like in spin-ice systems \cite{Lammert,Khomskii}. 
\begin{figure}[!htp]
\begin{center}
\includegraphics[width=6cm,keepaspectratio]{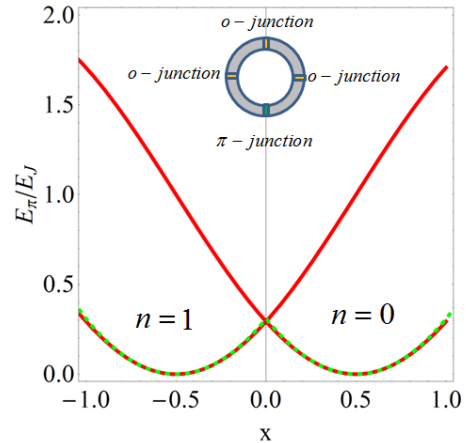}
\end{center}
\caption{(Color online) A  $\pi$-ring with four Josephson junctions is shown (three ordinary junctions and one  $\pi$-junction). The energy of the ring  is normalized by the Josephson energy  and is a function of the magnetic flux, $x=\Phi/\Phi_0$ , in units of the elementary flux quantum $\Phi_0$. The red lines are generated by Eq.(\ref{eq1}) for the Josephson cosine potential with zero or one vortex in the ring ($n=0$ or $n=1$, respectively). The green dashed line is the ground state energy that is described by Eq.(\ref{eq2}) with minima at $x=\pm1/2$.}
\label{fig1}   
\end{figure}
\\
We highlight superconducting rings for use as components of metamaterials due to their unique properties and also the spontaneous magnetizations that develop in  $\pi$-rings. In recent times these structures have become easier to make due to developments in fabrication technologies for molecular beam epitaxy, laser ablation and electron lithography techniques. These advancements in fabrication technologies allow one to design and to develop novel quantum devices with different structures built up of Josephson junctions and  even nanopatterned superconducting structures \cite{Cordoba}. The penetration of the magnetic field into a single ring with Josephson junctions, the simplest structure of which is known as a Superconducting Quantum Interferometer Device (SQUID), is characterized by the absorption of the magnetic flux quanta due to a flux quantization as it follows from gauge invariance.  This flux quantization forms a firm basis for a highly sensitive measurement of magnetic fields. Here, in the present paper, we propose a series of new devices which are a generalization of a single SQUID to their arrays.  Such an array allows one to make measurements not only of the DC magnetic field but also AC or microwave electromagnetic fields with a high precision. In an array there emerges an interaction between single Josephson vortices which are trapped with the magnetic field by the elementary rings or SQUIDs of the Josephson network structure. The way these vortices penetrate into it depends upon the shape of the structure and whether it has a plane or space geometry. 
\\
In the present work we consider simple networks of two-dimensional arrays of Josephson rings and calculate their basic vortex configurations.  Particular attention is given to arrays consisting of Josephson rings with a single  $\pi$-junction included ($\pi$-rings). The simplest Josephson structures of  $\pi$-rings are shown in the Fig.\ref{fig2} as linear arrays. In our method, we consider several examples of Josephson networks where the flux patterns may have some applications. The work focuses upon chains of rings with periodic boundary conditions.\\
Different two-dimensional Josephson arrays have already been investigated experimentally, in detail, by the means of the mutual inductance methods in \cite{Leemann,Rao}. The Josephson arrays containing a few cells have been studied in the framework of the resistivity-shunted-junction model \cite{Koelle} in order to understand how big the real disorder and frustrations of the lattice are. There are numerous experimental investigations in this field that were carried out in the past [e.g., \cite{Koelle,Orlando,Theron,Davidovic}]. In the present work, considering arrays of Josephson rings with $\pi$-junctions, we found the distribution of the vortices for the given values of magnetic field for both the ground and the excited states. The incorporation of an odd number of $\pi$-junctions into the rings creates a set of highly degenerate energy states that otherwise would only be possible with the application of a magnetic flux equal to half an elementary flux quantum. The $\pi$ phase shift induces a spontaneous current that can flow clockwise or counter-clockwise around the ring \cite{kusmartsev1992,Ortlepp}. In both isolated and coupled groups of  $\pi$-rings the magnetic fluxes tend towards one-half an elementary flux quantum as the large inductance limit is approached \cite{Ortlepp}. 
\begin{figure*}[!htp]
\begin{center}
\includegraphics[width=16cm,keepaspectratio]{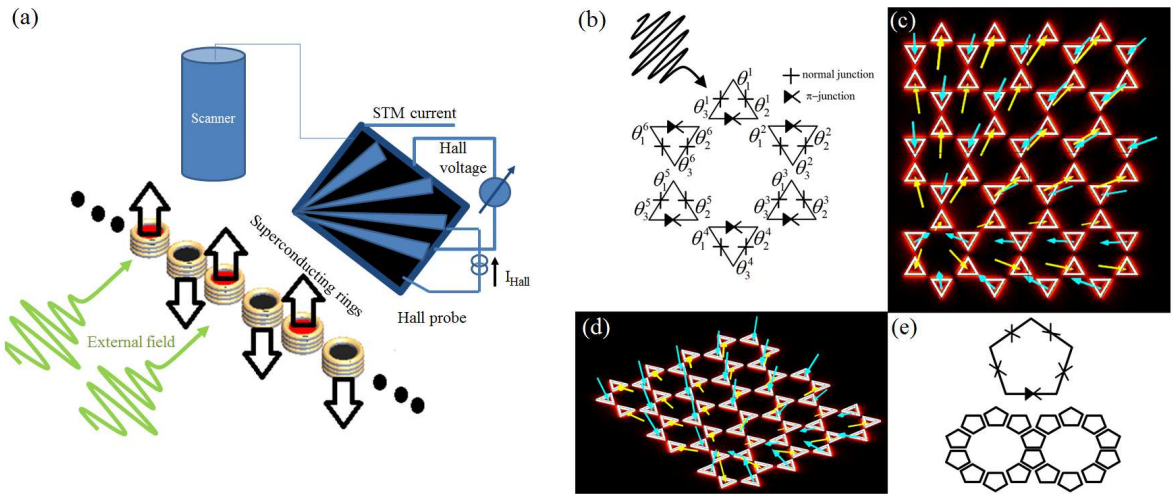}
\end{center}
\caption{(Color online) (a) The configurations of the magnetic moments of the simplest linear array of superconducting rings, whether they be Josephson  $\pi$-rings or more generally a superconducting ring biased in an external flux, can be sensitively determined by a scanning Hall probe microscope \cite{Davidovic}. (b) In this paper we examine the excited and ground state spontaneous magnetizations in chains of  $\pi$-rings. (c)-(d). The methods in this paper can easily be applied to two dimensional arrays such as the electronically isolated hexagonal lattice (as pictured), where the magnetic fluxes (yellow, positive orientation and blue, negative orientation arrows)  can exhibit a complicated arrangement dictated by frustration, edge effects and spatial separation. The $\pi$-rings can take on a range of geometries, with $N$ superconducting islands and Josephson junctions, e.g. (e) where we show a pentagonal ring and its inception as part of  a two dimensional cluster.}
\label{fig2}   
\end{figure*}

\section{Electronically isolated  $\pi$-rings}

The interaction of an assembly of closely spaced, electronically isolated $\pi$-rings leads to a rich and complicated energy balance. Fabrication techniques are sufficiently advanced for it to be legitimate to assume that all the junctions on the ring have the same coupling energy, $E_J=\hbar I_c\left(T\right)/2e$, where $I_c\left(T\right)$ is the temperature dependant junction critical current. Because of the flux quantization or the Aharonov-Bohm effect, the value of the flux is limited to $x=\Phi/\Phi_0<1/2$, i.e. the phase change over the junction is limited to the values $\theta_i-\theta_j\approx2\pi x/N_J<\pi/N_J$. This indicates that if the ring has many junctions, $N_J>>1$, that the change of the phases $\theta_i-\theta_j$ on a single junction $\left\langle ij\right\rangle$ of the ring may be chosen by an appropriate gauge transformation so that it does not vary very much. If there is a single $\pi$-junction in the ring then there is a current in the ground state \cite{kusmartsev,giles}. This supercurrent, $I_s$, is associated with the phase differences $\theta_i-\theta_j$, which occur due to magnetic flux and the $\pi$-shift in the summed phase difference. Therefore, for each $\pi$-ring an orbital moment arises. So then the arrays of the  $\pi$-junction rings can be considered as an array of orbital moments. That is, if we consider the case of a complicated multi-connected network of the interacting $\pi$-rings, the gauge symmetry dictates that the total magnetic flux   through each ring in this network must be quantized.  With a single $\pi$-junction the expression for the total energy of the ring has the form \cite{kusmartsev1992}: 
\be
E\left(x\right)=E_J\left(1-cos\left(\frac{2\pi n}{N}-\frac{2\pi x}{N}-\frac{\pi}{N}\right)\right)\label{eq1}.
\ee
This is the ground state energy for a single ring consisting of $N_J$ ordinary junctions, including the magnetic energy of the ring, the Josephson coupling energy, and with one $\pi$-junction, $N_\alpha=1$, i.e. $N=N_J+N_\alpha$. The last term in the cosine function, $-\pi/N$, stands for the existence of a single $\pi$-junction in the ring. In a ring there can be the presence of a vortex state $\left(n=1\right)$. The resulting energy potential can be seen in Fig.\ref{fig1}, where one can see that the cuspoidal shape accurately describes the low energy term for $\left|x\right|\leq1/2$,  
\be
E_\pi\left(x\right)=\frac{2\pi^2 E_J}{N^2}\left(\frac{1}{2}-\left|x\right|\right)^2\label{eq2}.
\ee
The zero flux cusp in the energy flux dependence corresponds to the instability leading to the formation of the spontaneous current and the orbital moment of the  $\pi$-ring. Hence, we obtain a model of the orbital moment formation using a very simple harmonic approximation \cite{kusmartsev2}. Here we assume that $E_J<0$. Then the ring has trapped a unit flux and a harmonic approximation can also be carried out around the new minima associated with the single trapped vortex in the ring. Without a trapped vortex, the energy minimum corresponds to the flux $x=1/2$. To describe the interaction of the  $\pi$-rings, 
\be
\epsilon=\frac{\Phi_0^2}{2L}\sum^{N_R}_{i=1} \gamma \left(\frac{1}{2}-\left|x_i\right|\right)^2 + x_i^2+\beta x_ix_{i+1},
\label{eq3}
\ee
where, $N_R$ is the number of $\pi$-rings and $x_i$ is the spontaneous flux in units of the elementary flux quantum $\left(\Phi_0=2.07 \times10^{-15}Wb\right)$. The self and mutual-inductances, $L$ and $M$ respectively, are taken to be of constant value for each ring and ring-ring interaction in an array. The rings are coupled through mutual inductance. The parameters $\gamma=2\pi L I_c/N^2 \Phi_0$ and $\beta=2M/L$ define the magnetic properties of the coupled rings. The self-inductance of each of the rings is $E_{self}=LI_s^2/2=\Phi^2/2L$, where $I_s$ is the current circulating around each ring and $\Phi=LI_s$. Note that for an $N$ junction ring, $\gamma$, is similar to the well-known parameter, $b=2\pi L I_c/\Phi_0$, that is used to define the inductance limit for the onset of spontaneous flux that is close to half a flux quantum \cite{Li, Tian, Li2}. It is worth noting that this parameter is sensitive to temperature fluctuations and applied magnetic fields. For example, random temperature fluctuations or the application of an electromagnetic field can change the system character and thus the effective dielectric constant of an array of $\pi$-rings, which may be a useful fact when designing metamaterials.

\begin{figure}[!htp]
\begin{center}
\includegraphics[width=8.5cm,keepaspectratio]{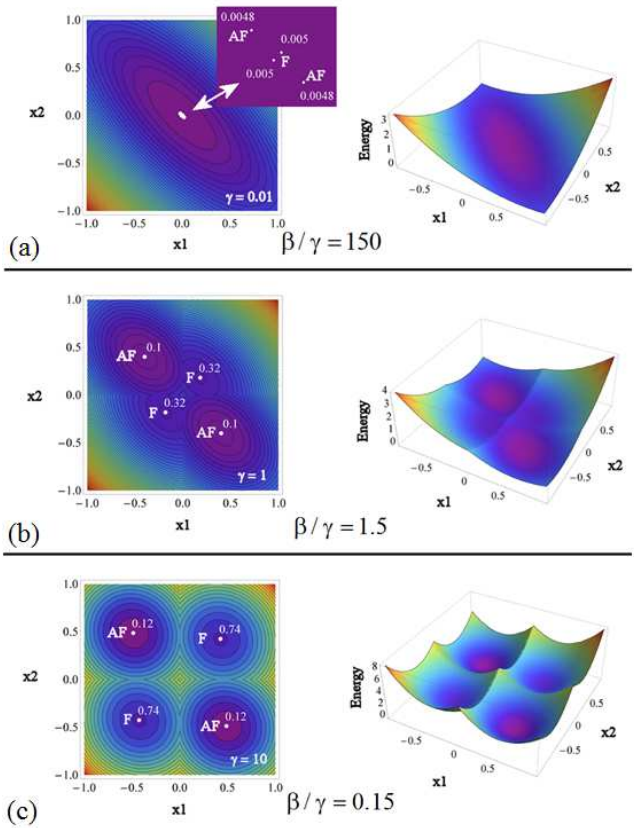}
\end{center}
\caption{(Color online)  The energy landscape as a function of the spontaneous fluxes, $x_1$ and $x_2$, for two coupled $\pi$-rings with $\beta=15$. $(a)$ A small value of $\gamma=0.01$ gives ground state energies that lie in very close proximity to one another (see the magnification). The white dots represent the energy minima, $F$ represents a ferromagnetic coupling and $AF$ an antiferromagnetic one. The numbers next to each white dot indicate the energy value in dimensionless units (see Eq. (\ref{eq3})). $(b)$ $\gamma=1$. $(c)$ $\gamma=10$. It can be seen that the lowest energy state occurs when there is an $AF$ configuration.}
\label{fig3}   
\end{figure}

\begin{figure}[!htp]
\begin{center}
\includegraphics[width=8cm,keepaspectratio]{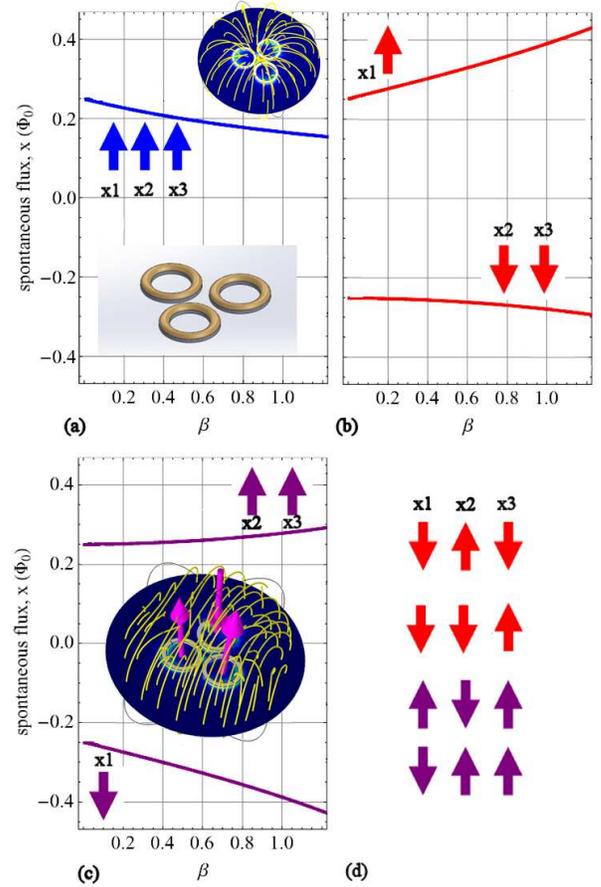}
\end{center}
\caption{(Color online)  Magnetic flux quanta with either up or down polarity in a system of three interacting $\pi$-rings (inset of $(a)$). The spontaneous flux is shown for a range of values of $\beta$ at $\gamma=1$. In $(a)$ a ferromagnetic state is illustrated. In $(b)$ and $(c)$ the frustration in the energy interaction of three rings causes a strain on the flux polarity of one of the rings due to the neighbouring rings influences. This energy balance results in there being no antiferromagnetic state when there is a triangular arrangement of the rings. The ferromagnetic state may be induced by an external magnetic field contribution in Eq. (\ref{eq1}).  In $(d)$ the other orientations of the magnetic flux, which are degenerate to those shown in $(b)$ and $(c)$, are depicted by the up and down arrows for each $x_i$ with the top two (red online) corresponding to $(b)$ and the bottom two to $(c)$ (purple online).}
\label{fig4}   
\end{figure}

\begin{figure*}[!htp]
\begin{center}
\includegraphics[width=13cm,keepaspectratio]{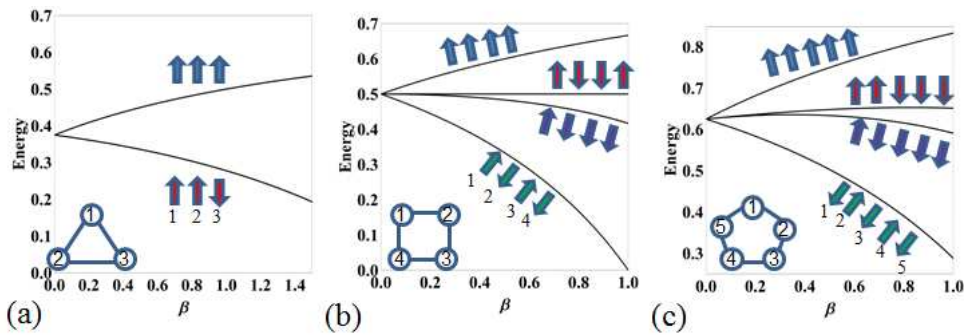}
\end{center}
\caption{(Color online) The energies according to Eq.(\ref{eq3}) and divided by $\Phi_0^2/2L$ for $(a)$ three, $(b)$ four and $(c)$ five $\pi$-rings for $\gamma=1$ and increasing $\beta$. In the left corner of each plot is the sketch of the geometrical arrangement of the rings, with each ring given an index, $i=1,...,N_R$. Next to each energy branch there are arrows depicting the polarity of the flux threading the rings: up for positive and down for negative. Each arrow is allocated a number which corresponds to the index of the ring it is representative of. In $(a)$ there are two energy branches: one for when the magnetic polarities in the rings are the same and the other when there are two polarities in opposite direction to the third. Likewise, in $(b)$ there are four degenerate energy branches. The highest energy branches are for the $F$ states. The next highest branch is when there are two neighboring rings with the same polarity. Then the next highest energies are those corresponding to three equal polarities in opposition to a fourth. The lowest energies are the $AF$ configuration. In $(c)$, when $\beta=1$, the configuration is $F$ at energy $0.83$. When the energy is $0.65$ there is a ten-fold degeneracy for combinations of the type $\uparrow\uparrow\downarrow\downarrow\downarrow$. There is another ten-fold degeneracy when there is one polarization opposite to the rest, e.g. at energy $0.59$ with $\beta=1$. And another for $\uparrow\downarrow\uparrow\downarrow\downarrow$  like combinations, e.g. at energy $0.29$ with $\beta=1$.}
\label{fig5}   
\end{figure*}

\begin{figure}[!htp]
\begin{center}
\includegraphics[width=8.5cm,keepaspectratio]{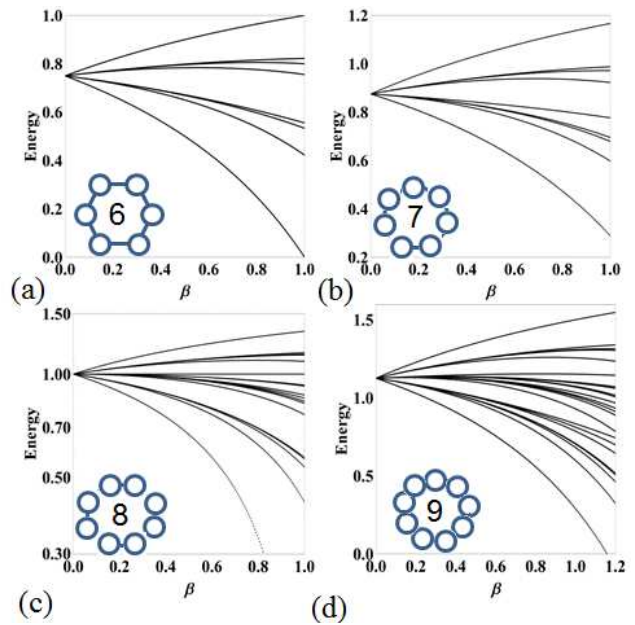}
\end{center}
\caption{(Color online) The energy against $\beta$ plots for six to nine rings when $\gamma=1$ are shown in $(a)$ to $(d)$. In $(a)$ there are eight energy branches, in $(b$) there are nine, in $(c)$ there are eighteen and in $(d)$ there are twenty-three. The $F$ states have the highest energies in all cases and correspond to the highest energy branches. The lowest energies are the $AF$ states in even numbers of rings and as close to an $AF$ state as is possible in the odd numbered rings (e.g. $\downarrow\uparrow\uparrow\downarrow\uparrow\downarrow\uparrow$, when $N_R=7$).}
\label{fig6}   
\end{figure}

\begin{figure}[!htp]
\begin{center}
\includegraphics[width=8.5cm,keepaspectratio]{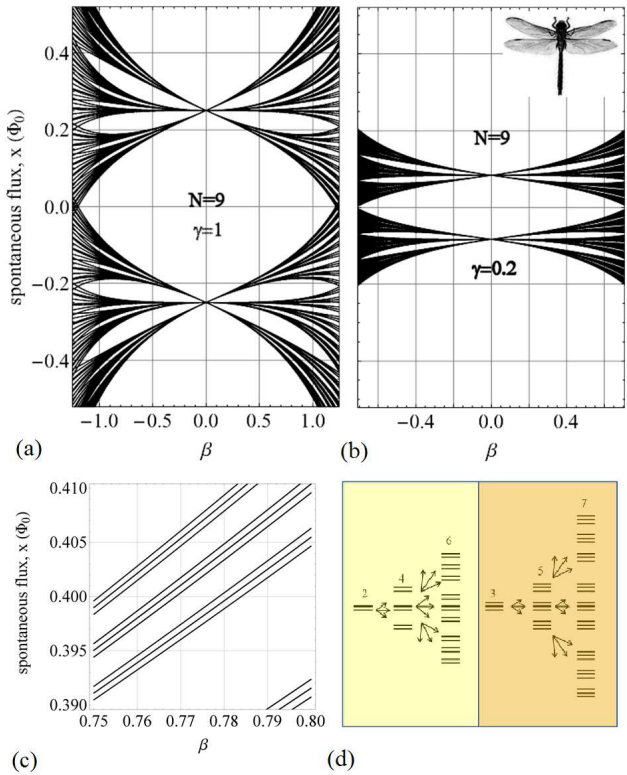}
\end{center}
\caption{(Color online) $(a)$ For nine $\pi$-rings, the spontaneous flux as a function of $\beta$ for values of $\gamma=1$ and $(b)$ $0.2$ is shown. There is a clear flux band structure that is interspaced by flux band gaps. The pattern seen above is most prominent when $N_R>6$ and is repeated as $N_R$ tends to infinity. In $(c)$ the ``sub-bands'' of (a) are magnified for part of the plot. The hierarchy to the development of this "dragonfly" fractal is described schematically in $(d)$. It demonstrates the band and gap structure that scales with the number of rings. For nine rings there are 27 scaled copies of the $N_R=3$ cell. In $(d)$ on the left is the band structure for even numbers of rings and on the right the band structure for odd numbers. The amount of rings is depicted by the number above each cluster. When there are two $\pi$-rings, the band has two branches and when there are three rings there are three. The bands of two and three rings are replicated as the size of the system is increased, but in clusters of three. Odd and even numbers of rings have the same type of band structure, although there does not appear to be an obvious scaling law. For odd numbers of rings there are $3^{(N_R-1)/2}$ branches of positive (or negative) polarity in total. For even numbers of rings there are $2\times 3^{(N_R-2)/2}$ branches of positive (or negative) polarities of flux. In both even and odd numbers of rings there are six main bands that scale as in $(d)$.}
\label{fig7}   
\end{figure}

In this model the limits are as follows. When $\beta/\gamma\rightarrow \infty$, the minima of the energy correspond to the zero flux state where $x_i\rightarrow 0$ (see Fig.\ref{fig3} (a) to see this situation in the flux contour plots, with $\beta/\gamma=150$).
At arbitrary values of $\gamma$ and $\beta$ there is a competition between these two parameters. Such a competition may lead to multi-stability and the formation of a fractal. The fractal emerges as $N_R\rightarrow \infty$, but in order to map its origin we see in Fig.\ref{fig3} $(b)$, where there are two coupled rings, that $\left|x_i\right| <1/2$ (i.e. $\beta/\gamma=1.5$). In the limit $\beta/\gamma\rightarrow 0$, the minima correspond with that of the equal flux state, where flux on all rings is equal to $x_i=\pm1/2$ (see Fig. \ref{fig3} $(c)$, where $x_i=\pm1/2$ configurations are shown for $\beta/\gamma=0.15$). The $\gamma$ parameter corresponds to the development of the long-range coherent state on each superconducting ring, while $\beta$ measures the strength of the coupling between the rings. Thus, when $\gamma\rightarrow 0$ and $\beta$ is fixed no supercurrent exists and $x_i=0$. When $\gamma\rightarrow \infty$ and $\beta\rightarrow0$, or is fixed, $x_i=\pm1/2$  and disorder happens. The final limiting case is for fixed $\gamma$ and $\beta\rightarrow\infty$, where antiferromagnetic ordering in the orientation of the magnetic flux ensues.
To find the energy minima, the first order partial derivatives of the system of rings are calculated ($\partial \epsilon_{\alpha}/\partial x_i=0$). In Fig.\ref{fig4}, the spontaneous fluxes in three  $\pi$-rings are shown.
Figure \ref{fig4} $(a)$ depicts the condition whereby all of the spontaneous fluxes are aligned ferromagnetically ($F$), $\uparrow\uparrow\uparrow$ or $\downarrow\downarrow \downarrow$. The value of $\gamma$ is set to one. The coupling parameter $\beta$ is restricted to be less than two (i.e. $M<L$). As $\beta$ increases, with $\gamma$ remaining constant, the absolute value of the spontaneous flux for a $F$ state decreases, whilst the energy level becomes higher. The $F$ states give the smallest values of $\left|x\right|$ , as can be seen by comparing Fig.\ref{fig4} $(a)$ with $(b)$ and $(c)$. Spontaneous fluxes, associated with the $\left\{\uparrow\downarrow\downarrow,\downarrow\uparrow\downarrow,\downarrow\downarrow\uparrow\right\}$ states occupy the same energy level and also occupy the same flux branches in Fig.\ref{fig4} $(b)$. Likewise, the $\left\{\uparrow\uparrow\downarrow,\downarrow\uparrow\uparrow,\uparrow\downarrow\uparrow\right\}$ states occur at the same level of energy minima and lie on the same spontaneous flux branches (see Fig.\ref{fig4} $(c)$). In Fig.\ref{fig4}, the spontaneous fluxes for a triangular arrangement of the rings are shown. The three rings have a frustration due to the closed boundary conditions causing an energy competition much like in a system of frustrated Ising spins \cite{kusmartsev3}. Frustrated arrays have highly degenerate ground states. Only two of the three spontaneous fluxes can align anti-parallel, so all possible mixed polarity configurations of the flux correspond to ground state energies. The ferromagnetic state arises with an excited energy level and may emerge in response to system temperature variations along with external radiation or fabrication deviations in the sizes of the rings.           The degenerate energy branches, as a function of  $\beta$ and $\gamma=1$, are illustrated in Fig.\ref{fig5} for three, four and five rings.
Taking four rings as an example, in the $F$ state the energy $\epsilon=0.64\Phi_0^2/2L$ at $\beta=0.8$, and the spontaneous flux, $x_i=\pm 0.179$. The configurations of the spontaneous fluxes, e.g. $\uparrow\uparrow\uparrow\downarrow,\downarrow\uparrow\uparrow\uparrow,\uparrow\downarrow\uparrow\uparrow,\uparrow\uparrow\downarrow\uparrow,\downarrow\downarrow\downarrow\uparrow etc$, which give the first excited degenerate energy state can be seen in Fig.\ref{fig5} $(b)$. In these orientations, $\epsilon=0.45\Phi_0^2/2L$. In Fig.\ref{fig5} $(b)$ the $\downarrow\downarrow\uparrow\uparrow,\uparrow\downarrow\downarrow\uparrow etc$ states have spontaneous flux orientation, $x_i=\pm 0.25$ and thus energy $\epsilon=0.5\Phi_0^2/2L$. The ground state energy occurs for an $AF$ state such as that shown in Fig. \ref{fig5} $(b)$: $\epsilon=0.17\Phi_0^2/2L$ and $x_i=\pm 0.417$. In Fig. \ref{fig5} we show the energies for a triangular, square and pentagonal arrangement of rings, with each ring positioned at one of the vertices. There are $2^{N_R}$ configurations of the spontaneous flux in each system, but one can see in Fig. \ref{fig5} that there is a high degree of energy degeneracy. The origin of this degeneracy is simple: there is one energy branch for $F$ states, one for $AF$ states, one for the case of one oppositely polarized ring to the rest, another for two oppositely polarized rings $etc$. Of course, the degree of separation between the oppositely polarized rings spawns a set of degenerate energy levels too. For example, in a system with six rings such as in Fig.\ref{fig6} $(a)$, the $\uparrow\uparrow\downarrow\downarrow\downarrow\downarrow$ and $\uparrow\downarrow\uparrow\downarrow\downarrow\downarrow$ configurations have $\epsilon=0.82\Phi_0^2/2L$ and $\epsilon=0.42\Phi_0^2/2L$, respectively, when $\beta=\gamma=1$.           
In Fig.\ref{fig6} one can see the beginning of the formation of energy bands. The interaction of larger numbers of $\pi$-rings gives rise to a spontaneous flux as a function of $\beta$. There are values of spontaneous flux that cannot be generated in a $N_R$ ring system. The signature of disallowed levels of magnetic flux can be seen for low values of $N_R$ (e.g. see Fig.\ref{fig4} for three rings) but does not become completely obvious until larger values of $N_R$. For example, the disallowed values of the spontaneous flux can be seen to be very pronounced in Fig.\ref{fig7}. Figure \ref{fig7} shows the branching of the spontaneous fluxes for nine coupled $\pi$-rings and hence the flux band structure of the system. For the nine ring system there is no completely $AF$ state (due to the periodic boundary conditions), as is the case for all odd numbers of $\pi$-rings. The gaps in Fig.\ref{fig7} form a pattern that resembles a butterfly, or perhaps more accurately dragonfly wings. This is due to the flux bands clustering into groups of three which themselves cluster into groups of three in a repeating pattern. The dragonfly fractal has an appearance that may remind one of the Hofstadter butterflies \cite{Hofstadter}.     
The $\pi$-rings are modelled on a two-dimensional lattice with each ring in close proximity.  The plot in Fig.\ref{fig7} has large gaps that visually appear like ``wings''.  There are fine flux bands that exist around the main wings. The bands of spontaneous flux found for the even numbers of $\pi$-rings, follow a pattern whereby the bands cluster into groups of three, which themselves cluster into groups of two, and so on. The exception is for two rings, for which there are two flux bands consisting of two branches (one band for positive polarity and the other for negative polarity of the spontaneous flux). Each flux branch is given its own value of spontaneous flux that is related to that of an individual $\pi$-ring. The result is a fractal nature. Our results allow us to determine a spectral picture in terms of the magnetic flux that is related to the number of rings in the system. Thus we have found a spectrum for every physically plausible value of $\beta$. There is a complicated structure of spontaneously generated fluxes that owe their existence to the formation of spontaneous persistent currents.       
\begin{figure}[!htp]
\begin{center}
\includegraphics[width=8.5cm,keepaspectratio]{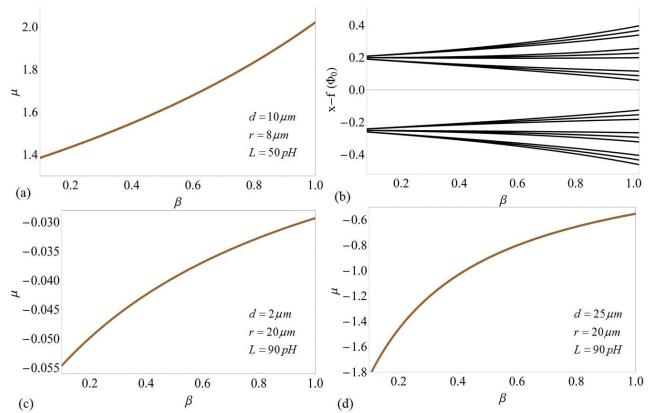}
\end{center}
\caption{(Color online) A magnetic flux of $f=0.05$ is applied to an array of five $\pi$-rings. Each ring contains a single $\pi$-junction and three ordinary junctions. In $(a)$, $(b)$ and $(d)$ the effective relative permeability of the array is shown as a function of $\beta$. In $(a)$ positive permeability for rings with parameters $L=50pH$, radii $r=8\mu m$ and $d=10\mu m$ is demonstrated. In $(c)$ and $(d)$ values of negative permeability are found by altering the geometry of the rings. The parameters are shown in the bottom left corner of each plot. The fractal structure of the flux as a function of $\beta$ is shown in $(b)$.}
\label{fig8}   
\end{figure}
\section{Superconducting metamaterial composed of $\pi$-rings}         
Metamaterials are artificially fabricated in order to elicit unusual electromagnetic responses and negative permeability materials can be created through the collective interaction of the metamolecules or atoms with one another and the applied electromagnetic fields. The electric permittivity, $\epsilon_p$, and the magnetic permeability, $\mu$, are properties of the material that describe its electromagnetic response. In physical terms, the permeability describes how the magnetic fields affect the material, but also how the material influences the magnetic field. Small magnetic fields are introduced to the $\pi$-ring structures in order to demonstrate the manipulation and emergence of negative values of the permeability. The applied magnetic field is denoted by $H$ and in $SI$ units the magnetic permeability is $\mu=1+(M_g/H)$. By definition, the magnetic field strength, $H$, is given in terms of the magnetic field density $B$ and the magnetization $M_g$ as $H=(B/\mu_0)-M_g$. The magnetization is defined as $M=m/V=j S/V$, where $m$ is the magnetic moment, $V$ is the volume, $j$ is the orbital current and $S$ is the cross-sectional area threaded by the magnetic flux. We introduce $f$ as our dimensionless externally applied flux, $f=\Phi_{ext}/\Phi_0$ with the magnetic flux penetrating the rings $\Phi_{ext}=BS$. Thus, the magnetic permeability can be written as,
\be
\mu=\frac{f d \Phi_0}{fd\Phi_0-\mu_0jS}\label{eq4},
\ee      
where $d$ is the thickness of a $\pi$-ring. We find the orbital current by differentiating the energy with respect to $x$, $j=-\partial\epsilon/\partial x$ \cite{kusmartsev1992}, resulting in,
\be
j_i=\frac{\Phi_0}{2L}\left[Y-\left(\gamma+2\right)x_i+\left(2+2\beta\right)f-Z\right]\label{eq5},
\ee 
where \be Y=\gamma\left(1-2f\right)sgn\left(x_i\right)\ee 
and \be Z=\beta\left(x_{i+1}+x_{i-1}\right).\ee
The ring can be thought of as a closed loop of wire, with the thickness of the $\pi$-ring, $d$, taken to be the diameter of the wire. To illustrate the $\pi$-ring array as a positive or negative permeability metamaterial, we give the example of five rings arranged in a pentagonal lattice in Fig.\ref{fig8}. In Fig.\ref{fig8} $(a)$ each ring has a self-inductance of $L=50pH$, radius $r=8\mu m$ and a thickness of $d=10\mu m$. In all the plots in Fig.\ref{fig8} the array is subjected to a perpendicular magnetic field, that gives rise to an external flux of $f=0.05$. This produces a shift in the flux fractal pattern as shown in Fig.\ref{fig8} $(b)$. Changing the size of the ring radically changes the effective permeability.      
In Fig.\ref{fig8} (c), the inductance of each of the rings is $90pH$ with radii of $20\mu m$ and thicknesses of $2\mu m$. For this ring geometry we find that the system exhibits negative permeability as a function of the parameter $\beta$. Increasing the thickness of the ring can increase the magnitude of this negative permeability, as in Fig.\ref{fig8} $(d)$. Thus, here there is a possible mechanism to tune the permeability of an array of $\pi$-rings that are subjected to small static fields. The parameters in Fig.\ref{fig8} are similar to those found in experiments (e.g. \cite{Kirtley2}). 
\begin{figure}[!htp]
\begin{center}
\includegraphics[width=8cm,keepaspectratio]{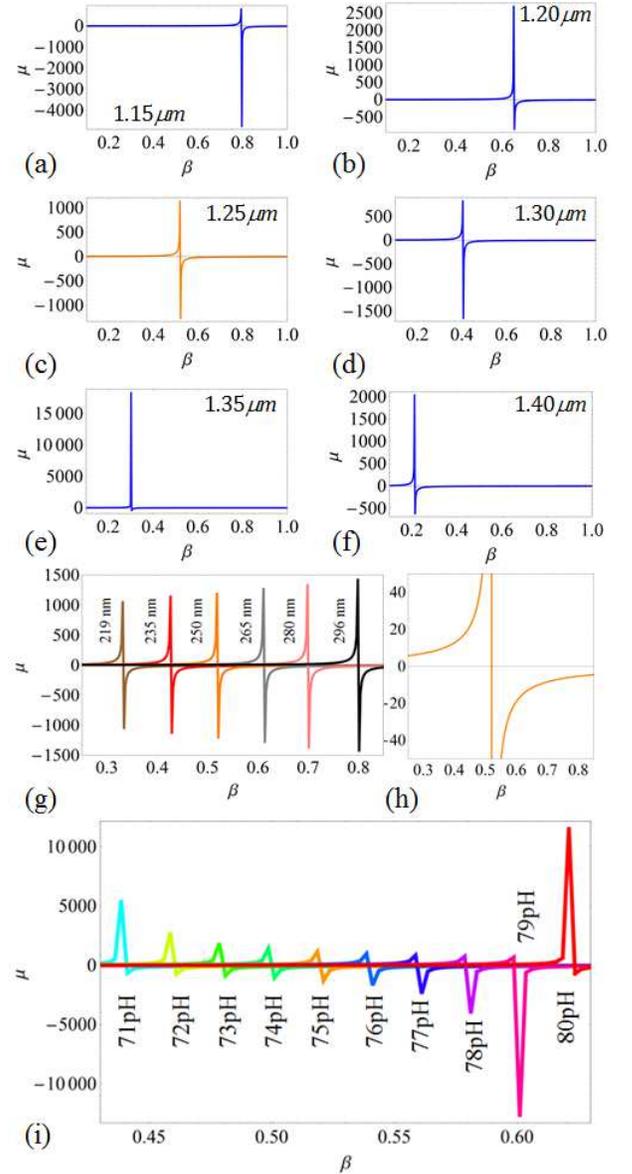}
\end{center}
\caption{A tunable level of effective relative permeability where inductance $L$, thickness $d$ and radius $r$ can strongly influence the magnetic response. In $(a)$ to $(e)$ the values of $L$ and $d$ are taken as constant at $75pH$ and $250nm$, respectively. As the radius $r$ of the ring changes from $1.15\mu m$ in $(a)$ to $1.4\mu m$ in $(f)$ there emerge a series of peaks and switching events in the effective permeability of the $\pi$-ring array. In $(g)$,  $L=75pH$ and $r=1.25\mu m$ and the radius is changed from $219nm$ to $296nm$. A magnification of the case of $L=75pH$, $r=1.25\mu m$ and $d=250nm$ (orange curves throughout) is shown in $(h)$. In $(i)$ the inductance changes from $71$ to $80pH$ with $r=1.25\mu m$ and $d=250nm$.    }
\label{fig9}   
\end{figure} 
The separation between the $\pi$-rings, and their size distribution holds the key to finding complicated permeability dependences and hence a strong reliance on the value of $\beta$. Figure \ref{fig8} showed that there was either a positive or negative permeability over all values of $\beta$ in the physically valid range of $\beta=0$ to $1$ for the aforementioned parameters used. We now demonstrate that with carefully chosen  system characteristics that the sign and amplitude of $\mu$ can change within this range. In Fig.\ref{fig9} $(a)$ there is a positive permeability with $\beta\leq 0.8$, but after this value the system has negative permeability. This feature is typical of all the situations in Fig. \ref{fig9}. The sign cross-over is marked by a large permeability peak. The self inductance is contained within the $\beta$ term which is a ratio of $L$ with respect to the mutual inductance of the system. Figure \ref{fig9}  $(a)$ to $(f)$ illustrate the sensitivity to changes in the radius of the rings with $L$ and $d$ chosen to be $75pH$  and $250nm$, respectively. Even small changes to the radii of the rings can alter the permeability, and it can be seen that by increasing the size of $r$ that $\beta$ becomes smaller. It is also noteworthy that the peaks of the Lorenz-like curves in Fig.\ref{fig9} can be maximized or reduced by slight variations of the radius which controls the value of $\beta$. For example, in Fig \ref{fig9} $(c)$ either side of $\beta\approx0.5$ (with $r=1.25 \mu m$) the positive and negative permeability are of comparable magnitude, i.e. the maximum $\left|\mu^{+}\right|\approx\left|\mu^{-}\right|$. However, in Fig \ref{fig9} $(b)$, $\left|\mu^{+}\right|_{max}>>\left|\mu^{-}\right|_{max}$ with $r=1.20 \mu m$, whereas the opposite is true in $(d)$ for $r=1.30 \mu m$. Note that changes in the radius alter the position of the peaks with respect to $\beta$. Figure \ref{fig9} $(g)$ illustrates how the thickness of the rings also shifts the peaks and here we have chosen to show the cases where $\left|\mu^{+}\right|\approx\left|\mu^{-}\right|$. Where these equal magnitude positive and negative permeabilities occur are ``pivotal'' positions and from $(g)$ one can see that for these particular $\left|\mu^{+/-}\right|$ an increase in permeability as $r$ is increased from $r=219.146nm$ (rounded to $219nm$ in the figure) to $296.046nm$ happens. Around these pivot positions the magnitude of the permeability fluctuates between having high positive values to very low negative ones (similar to the situation of Fig.\ref{fig9} $(e)$, where $\mu^{+}>15000$). This is also demonstrated as $L$ increases from $71pH$ to $80pH$ in Fig.\ref{fig9} $(i)$. These situations arise as one tailors the $\pi-ring$ array between diamagnetism and a paramagnetic Meissner state \cite{kusmartsev1992b,khomskii1994}.   

\section{Discussion}
In earlier works the orbital glasses and the paramagnetic Meissner effect were elaborated upon in the context of $\pi$-rings in granular superconductors \cite{kusmartsev1991,kusmartsev1992,kusmartsev1992b,FVKexp} to explain experimental anomalies in the Meissner effect. Interestingly Josephson coupling was seen between the copper oxide layers in BISCCO single crystals that behaved like a linear chain of Josephson junctions \cite{kleiner}.The currents spontaneously created in the $\pi$-ring loops are dependent upon dissipation in the network of Josephson junctions leading to an increase or decrease in the orbital currents \cite{kusmartsev1992}. The mutual inductance forms a coupling between the rings which can increase or decrease as a function of proximity and so influence the level of the orbital currents. Magnetism has even been previously found to coexist with superconductivity in the absence of the Meissner effect \cite{kusmartsev1992,kusmartsev1992b,FVKexp}. Thus, the $\pi$-ring array can have an amazing diversity of phase transitions and magnetic hysteresis pathways \cite{FVKexp} which we have found to lead to the possibility of a fractal structure in the magnetic flux. In the current work, the addition of a small magnetic field has been shown to produce remarkably complex changes between conditions for observing positive or negative permeabilities. The situations found in small magnetic fields experimentally in \cite{FVKexp} were also strongly influenced by temperature and this will be an area of future work for the fractal metamaterials discussed here. In this work a simplified model has been developed that has allowed us to characterize the spontaneous fluxes that develop in an array of $\pi$-rings.  It is interesting to note that in the work of Geim and co-workers \cite{geim} on the experimental verification of the paramagnetic Meissner effect for small superconductors, that changes in the radius and thickness of their superconducting disks altered the sign of the Meissner effect. A thinner disk resulted in a smaller field requirement for the paramagnetic Meissner state to become larger. In the fractal metamaterial, ring geometry also plays an important role, with the radius and nano-thickness dictating the size of the permeability and also the value of $\beta$ in a small external field. Recent advances in Josephson junction coupling have enabled the massive increase in the inductances of the circuitry through the addition of kinetic inductance to produce superinductors \cite{Masluk,Bell}. The $\pi$-ring arrays could be connected to one another through thin superconducting lines so that the total inductance becomes a sum of both the geometrical and kinetic inductances. In this way inductance can become three orders of magnitude higher \cite{Masluk,Bell}. Thus, ring geometries can also be created that have tunable inductances. 

We described the properties of a new fractal metamaterial that is consisting of interacting superconducting rings containing $\pi$-junctions. However, quantum confinement effects are associated with generic metamaterial systems and lead to novel properties amongst the ``meta-atoms''. The fractal metamaterial can also be found for structures such as those composed of nano-magnets (such as those in Ref. \cite{forrester2007}). Indeed, small linear chains described in terms of the Ginzburg-Landau model are well suited to finding novel spectroscopic signatures in the guise of fractals \cite{kuerten}. These kinds of systems can be described in a similar fashion to the $\pi$-rings and other metamaterials. For example, for two-dimensional arrays of conically shaped nano-pillars, like those fabricated on the nanoscale by Grigorenko and co-workers \cite{Grigorenko}, there can be the existence of a fractal spectrum. Small diameter cylinders or cones behave as if they have a single magnetic domain. The total distribution of magnetic moments in the molecular ring of ``artificial atoms'' is determined by dipole-dipole or exchange interactions. A chain of cylinders or cones with induced magnetism may be described by, 
\be F=-J\sum_{\left\langle i,j\right\rangle}\textbf{M}_i\textbf{M}_j+\sum_i\left(\frac{a}{2}\textbf{M}_i^2+\frac{b}{4}\textbf{M}_i^4+\textbf{H}\textbf{M}_i.
\label{eq8} 
\right)
\ee
Equation (\ref{eq8}) assumes a dipole-dipole coupling constant $J$ that describes the interaction of the meta-atoms. The external magnetic field is \textbf{H} and $a=a^{'}\left(T-T_c\right)$ and $b$ are phemonological constants. The magnetization is described by $\textbf{M}$, which for cones or cylinders can be taken in the reduced form to be $\textbf{M}=\left(M_x,M_y,M_z\right)=\left(0,0,M_s z_i\right)$. Following \cite{kuerten} and substituting for \textbf{M}, Eq.(\ref{eq8}) becomes,
\be
F=\left(\frac{a}{2}M_s^2-JM_s^2\right)\sum_{i}z_i^2+\frac{b}{4}M_s^4\sum_{i}z_i^4+\psi+\zeta
\label{eq9},
\ee 
where,     
\be \psi=\frac{JM_s^2}{2}\sum_{\left\langle i, j\right\rangle}\left(z_i-z_j\right)^2,\ee
and
\be \zeta=-HM_s\sum_{i}z_i.\ee
We now introduce some alternative notation in order to make the dimensionless form of the free energy: $\alpha=2-\left(a/J\right)$; $z_i=\sqrt{{\alpha J}/{bM_s^2}}x_i$ and $h=H \sqrt{b/\alpha J^3}$. In this way we now obtain a generic form for metamaterials that are composed of elements that have have an elongated magnetic structure.
\be
\textbf{F}\left(x_1,...,x_k\right)=\sum^{k}_{i=1}\frac{1}{2}\left(x_i-x_{i-1}\right)^2+\frac{\alpha}{4}\left(x_i^2-1\right)^2-hx_i.
\label{eq12}
\ee   
In the above $x_i$ defines the orientation of the magnetization in the i-th meta-atom and $M_s$ is an effective saturation magnetization. The energy $F$ has been scaled by $\textbf{F}=Fb/\alpha J^2$. This form of equation is similar to those employed for analyzing the coupled $\pi$-rings and will in fact also lead to fractal formation. For an in depth discussion of its use for a system of ellipsoidal nanomagnets see Ref.\cite{kuerten}. The point is that the fractal structures should be obtainable for a number of different kinds of magnetic structures, with complicated metastable energy states spawning the existence of these fractal clusters. So we have demonstrated the possible existence of a novel type of magnetic flux structure in $\pi$-ring arrays. It is also very interesting to note that ``splinter'' vortices have been experimentally found and to originate in a linear array of 0 and $\pi$-junctions  \cite{mints1,mints2}. It may be that arrays of $\pi$-junctions can also be created to discover experimental evidence for new vortex -anti-vortex structures and glasses (e.g. \cite{forrester}).

\section{Summary and Conclusions}

We have analyzed systems of Josephson $\pi$-rings that individually contain any number of normal Josephson junctions and one $\pi$-junction. These $\pi$-rings generate spontaneous magnetizations in the absence of a magnetic field. The number of rings in the system gives a distinctive spectral definition to the phase distribution as a function of the coupling parameter and the inductances of the system. The characteristics of odd and even numbers of rings gives slightly differing fractal patterns: even numbers of rings have an $AF$ state, whilst odd numbers do not. The results contained in this analysis may be of interest for applications such as the creation of reversible logic devices, quantum computation \cite{Feofanov}, parametric amplifiers and various metamaterial designs exploiting the low loss, negative permeability states of the system. The supercurrent in a single ring can flow in a clockwise direction, resulting in a downward magnetic moment. Flowing in the opposite direction (anticlockwise) the moment arises with an opposite polarity. The rings analyzed here are isolated from one another (no electronic coupling) and are thus coupled magnetically through mutual inductance. The application of small magnetic bias fields further manipulate the spontaneous currents and move the system from the ground state into a metastable state. Upon examining the ground states of these superconducting rings we have discovered the possibility of fractal patterns that describe the magnitude and the orientations of these fluxes. These appear as dragonfly wing fractals. Du $et$ $al$ have shown that SQUID arrays can be stable and bi-stable metamaterials \cite{Du} and Rakhmanov $et$ $al$ have developed quantum metamaterials \cite{rak}. We propose the $\pi$-ring array as a further enhancement upon this. Our description of the interactions between rings gives the ability to predict the behavior of small and large arrays that generate their own controllable interaction through the remarkable spontaneous generation of flux. The influence of external radiation on a superconductor array \cite{loo} can change the spectral pattern of the system, making the possibility for using these systems as sensitive detectors of electromagnetic fields. The analysis demonstrates the possible existence of a spectrum of spontaneous fluxes and hence the propensity for hysteresis paths between the multitude of flux-states to exist. 
\\       
\begin{acknowledgments}
This work has been supported by the European Science Foundation (ESF) in the framework of the network program ``Arrays of Quantum Dots and Josephson Junctions'' and the EPSRC KTA grant - ``Developing prototypes and a commercial strategy for nanoblade technology''.
\end{acknowledgments}

\end{document}